\documentclass[twocolumn]{aastex631}

\usepackage{fp}

\newcommand{\eviGALONE}{0.023}
\newcommand{\eviGALONEerr}{0.012}

\newcommand{\VbandextinctionGALONE}{0.045}

\newcommand{\IbandextinctionGALONE}{0.028}

\newcommand{\JbandextinctionGALONE}{0.014}

\newcommand{\FourFiveextinctionGALONE}{0.001} 

\newcommand{\ordinaryLSQval}{-0.50}
\newcommand{\ordinaryLSQerr}{0.12}
\newcommand{\weightedLSQval}{-0.56}
\newcommand{\weightedLSQerr}{0.11}
\newcommand{\invertedLSQval}{-1.06}
\newcommand{\invertedLSQerr}{0.18}
\newcommand{\Demingval}{-1.12}
\newcommand{\Demingerr}{0.25}
\newcommand{\ODRval}{-0.99}
\newcommand{\ODRerr}{0.16}

\revised{\today}

\graphicspath{{./}{figures/}}

\begin{document}
\title{Coordinated JWST Imaging of Three Distance Indicators in a SN Host Galaxy and an Estimate of the TRGB Color Dependence}

\correspondingauthor{Taylor J. Hoyt}
\email{taylorjhoyt@gmail.com}

\author[0000-0001-9664-0560]{Taylor J. Hoyt}
\affiliation{Physics Division, Lawrence Berkeley National Lab, 1 Cyclotron Road, Berkeley, CA 94720, USA}

\author{In Sung Jang}
\affiliation{Department of Astronomy \& Astrophysics, University of Chicago, 5640 South Ellis Avenue, Chicago, IL 60637}

\author{Wendy L. Freedman}
\affiliation{Department of Astronomy \& Astrophysics and Kavli Institute for Cosmological Physics, University of Chicago, 5640 South Ellis Avenue, Chicago, IL 60637}

\author{Barry F. Madore}
\affiliation{The Observatories of the Carnegie Institute of Washington, 813 Santa Barbara St, Pasadena, CA 91101}
\affiliation{Department of Astronomy \& Astrophysics, University of Chicago, 5640 South Ellis Avenue, Chicago, IL 60637}

\author{Abigail J. Lee}
\affiliation{Department of Astronomy \& Astrophysics, University of Chicago, 5640 South Ellis Avenue, Chicago, IL 60637}

\author{Kayla A. Owens}
\affiliation{Department of Astronomy \& Astrophysics, University of Chicago, 5640 South Ellis Avenue, Chicago, IL 60637}

\begin{abstract}
{Boasting a 6.5m mirror in space}, JWST {can increase} by several times {the number of supernovae (SNe) to which a redshift-independent distance has been measured with a precision distance indicator (e.g., TRGB or Cepheids); the limited number of such SN calibrators currently dominates the uncertainty budget in distance ladder Hubble constant ($H_0$) experiments.} 
{JWST/NIRCAM imaging of the Virgo Cluster galaxy NGC~4536 is used here to preview JWST program GO-1995, which aims to measure $H_0$ using three stellar distance indicators (Cepheids, TRGB, JAGB/carbon stars). 
Each population of distance indicator was here successfully detected--with sufficiently large number statistics, well-measured fluxes, and characteristic distributions consistent with ingoing expectations--so as to confirm that we can acquire distances from each method precise to about 0.05~mag (statistical uncertainty only).}
{We leverage overlapping HST imaging to identify TRGB stars, cross-match them with the JWST photometry, and present a preliminary constraint on the slope of the TRGB's F115W-($\mathrm{F115W}\!-\!\mathrm{F444W}$) relation equal to $-0.99 \pm 0.16$~mag/mag. This slope is consistent with prior slope measurements in the similar 2MASS $J$ band, as well as with predictions from the BASTI isochrone suite.}
{We use the new TRGB slope estimate to flatten the two-dimensional TRGB feature and measure a (blinded) TRGB distance relative to a set of fiducial TRGB colors, intended to represent the absolute fiducial calibrations expected from geometric anchors such as NGC~4258 and the Magellanic Clouds.}
{In doing so, we empirically demonstrate that the TRGB can be used as a standardizable candle at the IR wavelengths accessible with JWST.}
\end{abstract}

\keywords{cosmology: distance scale --- galaxies: individual (NGC~4536)
-- galaxies: stellar content --- stars: Population II}

\section{Introduction} \label{sec:intro}

{The trajectory of paradigm-shifting improvements in the realm of extragalactic distance measurement is closely tied to similar breakthroughs in observational capabilities. From the initial discovery of the expanding universe made possible by the 100'' telescope at Mt. Wilson \citep{Hubble_1929}, to the adoption of red-sensitive plates \citep[e.g.,][]{Baade_1944}, the development of CCD cameras \citep[e.g.,][]{Mould_1986, Freedman_1988_cephs, Freedman_1988}, and the launch of space telescopes \citep[e.g.,][]{Freedman_2001, Freedman_2012, Riess_2009, Riess_2016, Riess_2022}.}
{Now, JWST has been successfully launched, commissioned, and is in full science operation \citep{Rigby_2023, Gardner_2023}. With a 6.5m mirror diameter and newer IR detector technology than its predecessors, JWST has already begun rapidly improving the precision and accuracy of extragalactic distance measurements \citep{Lee_2024, Riess_2024, Anand_2024_shoes}.}

\subsection{The Hubble Constant}
The universe's present-day expansion rate, or the Hubble constant, $H_0$, continues to prove challenging to accurately measure via direct, astrophysical means such as the classical distance ladder. {After the Key Project resolved the factor of two debate using the Hubble Space Telescope (HST) and introduced comprehensive error budgeting into the distance scale \citep{Freedman_2001}, interest in measurement of $H_0$ was reignited because of the leverage it provides in constraining the dark energy equation of state when combined with cosmic microwave background (CMB) measurements \citep{Hu_2005}.}

{A departure from this initial goal of measuring $w$ began with claims of a ``Hubble Tension,'' which is a disagreement between local or direct measurements of $H_0$ (e.g., Cepheids and SNe) and those estimates that are tied to high-redshift observables (e.g., CMB or BBN+BAO) then extrapolated to the present day. The latest evidence from Cepheids-SNe suggests that the Hubble Tension has reached $5\sigma$ significance \citep{Riess_2022} and that new physics beyond the standard model must be the reason for it. However, the Cepheid-SN local measurement dominating the claim \citep{Riess_2022} may not fully account for some uncertainties that were demonstrated to have been previously underestimated \citep{Freedman_2023}.}

\subsection{JWST-GO-1995: Is the Tension in the Hubble Constant Real?}

{To address the nature of the Hubble Tension, a JWST program (PI: Freedman, co-PI; Madore, GO-1995) aims to assess how well the facility can measure extragalactic distances by comparing relative distances estimated to the same host galaxies using three different distance indicators}.
The program {is designed to acquire JWST/NIRCAM imaging of ten galaxies that have hosted Type Ia supernovae (SNe~Ia) as well as of three fields in the megamaser host galaxy NGC~4258, to which a 1.5\% trigonometric distance has been estimated \citep{Herrnstein_1999, Humphreys_2008, Humphreys_2013, Reid_2019}. These calibrator galaxies play the key role as the absolute anchors of the SN~Ia distance scale, and any subsequent SN-based derivation of $H_0$.}

{In each pointing, imaging was acquired so as to capture at the same time three distance indicators}: the classical Cepheid variables, the Tip of the Red Giant Branch (TRGB), and the predominantly carbon-enhanced population of red Asymptotic Giant Branch (AGB) stars that are found to populate a horizontal plume in the J-band color-magnitude diagram (referred to as JAGB). Each of the three methods can be naturally categorized by the galactic structures in which their respective stellar populations are optimally measured: the young, massive Cepheids in spiral arms and/or thin disks, intermediate-age JAGB stars in the thick, extended disk, and the old TRGB in the stellar halo (see \autoref{fig:pointings}).

The program {was} blinded {in terms of both the intra-sample relative distances as well as the sample-wide absolute zero point. This is done by injecting a random offset, different for each of the thirteen target fields, between $-0.2$ and $+0.2$~mag into each field's photometric catalog, then deleting all traces of the original magnitudes.}
After unblinding, each method's calibrated distance scale will provide a {new} measure of the Hubble constant and, should the cross-comparison reveal minimal systematic differences, the joint distance constraints can be leveraged for {a combined determination of the Hubble constant with improved uncertainties over any one method's constraint.}

{In this paper, each of the three methods will be showcased using JWST/NIRCAM observations acquired of the Virgo Cluster galaxy NGC~4536--with an outsize focus placed on the TRGB and its color dependence. We briefly introduce each method here and emphasize recent developments relevant to the infrared imaging considered here. For a comprehensive review of each method and of the 
motivation for program JWST-GO-1995, see \citet{Freedman_2023}.}

\subsection{Cepheid Variable Stars}

{Cepheid variables have long been the backbone of modern observational cosmology via application of the ``Leavitt Law,'' or the universal power-law relationship between a Cepheid's pulsation period and its phase-averaged luminosity \citep{Leavitt_1908, Leavitt_1912}. The Cepheid Leavitt Law was used to discover the expanding universe \citep{Hubble_1929}, and then revised via disambiguation between the classical and Pop II Cepheids, which populate separate period-luminosity relations \citep{Baade_1963}. It underpinned the factor of two debate over the age of the universe \citep[see, e.g.,][]{Sandage_1970, deVac_1978}, as well as its resolution, which began with some of the earliest observations of Cepheids with CCDs \citep[e.g.,][]{Freedman_1988_cephs, Freedman_1990} and finished with the HST Key Project \citep[HST-KP,][]{Freedman_2001}.}

{The HST-KP techniques were further developed and broadened in scope for use} with near-infrared (NIR) detectors on board HST such as NICMOS \citep{Macri_2001, Riess_2009} and WFC3/IR \citep{Riess_2016, Riess_2022}. {With Spitzer, extinction uncertainties could be reduced to negligible levels \citep{Freedman_2012}. This led to the discovery of a bandpass ($\lambda_{eff} = 4.5~\mu$m) in which a CO band-head could be observed and provide a means of acquiring direct abundance measurements of Cepheids from photometry alone \citep{Scowcroft_2011, Monson_2012, Scowcroft_2016}, at nearly the same precision as spectroscopy ($\sigma \simeq 0.2$~dex). The CO-band head has recently been directly confirmed with spectroscopic measurements of Cepheids \citep{Hamer_2023}.}

{JWST can directly build on the pioneering findings provided by Spitzer, which was limited to the nearby $\sim 100$~kpc universe. JWST increases the reach of 4.5 micron Cepheid measurements to the $~10$~Mpc universe and potentially beyond, thanks to its considerably better image quality--predominantly a function of the eight times increase in mirror size, though newer generation IR imaging technology also plays a major role. The measurement of Cepheid fluxes at 4.5 microns and the potential to directly measure Cepheid metallicities was considered in planning the JWST-GO-1995 observations of the most nearby targets (e.g., M101 and NGC~4258). For the remainder of the targets we chose to pair F115W with the F356W 3.6 micron band as it is the most sensitive of the NIRCAM/LW standard wide bandpasses. This excludes NGC~4536 and NGC~7250, which were imaged in F444W before we made the decision to switch to F356W for distant targets.}

\subsection{Horizontal ``J'' Branch of Carbon-enhanced Asymptotic Giant Branch Stars (JAGB)}

{2MASS observations of the LMC revealed a distinct branch of stars pulling off the narrow oxygen-rich asymptotic giant branch \citep[AGB][]{Nikolaev_2000}. \citet{Weinberg_2001} used that branch of stars to present a constraint on the geometry of the LMC that was consistent with prior determinations derived from established distance indicators (e.g., Cepheids), demonstrating that this branch of AGB stars was capable of providing precise distances when observed in the IR.}

{\citet{Madore_2020_jagb} updated the \citet{Weinberg_2001} findings by identifying from an expansive compilation of IR CMDs of the LMC \citep{Macri_2015, Hoyt_2018} that this branch of AGB stars became almost completely horizontal when observed in the J-band CMD. \citet{Ripoche_2020} found the same in 2MASS observations of the LMC and IR observations plus Gaia DR2 parallaxes of Milky Way stars. \citet{Freedman_2020_jagb} used this ``JAGB'' method to determine distances to 13 Local Group galaxies, comparing against the $I$-band TRGB and finding a total dispersion of 0.07~mag, thereby establishing the JAGB as a precision extragalactic distance indicator. The method has been vetted and refined upon by several independent groups, across different host galaxies, and tested with various approaches to the actual distance measurement methodology \citep{Lee_2021,Lee_2024_fourstar,Parada_2021,Parada_2023,Zgirski_2021}.}

\subsection{The Tip of the Red Giant Branch}
{When observed at the near-infrared (NIR) wavelengths at which a TRGB star's SED peaks, the TRGB brightness is strongly (and linearly) correlated with the metallicities/colors of coeval TRGB stars \citep{Salaris_1998, Salaris_2005, Bellazzini_2004, Bellazzini_2008, Madore_2018, Madore_2020, Durbin_2020}, contrary to its metallicity/color-insensitive (with a small, near-zero residual variance observed for metal-poor TRGB stars) manifestation when observed with bands that have effective wavelengths between 800 and 900~nm \citep{Lee_1993, Bellazzini_2001, Jang_2017_color, McQuinn_2019, Freedman_2021, Hoyt_2023, Anand_2024_shoes}. As a result, calibration of the TRGB's color dependence becomes the most important step to measuring precise and accurate TRGB distances in the NIR.} 

{There is general consensus in the literature that the shape of the NIR TRGB's color dependence is linear \citep{Valenti_2004, Bellazzini_2008, Serenelli_2017, Madore_2018, Durbin_2020}.\footnote{{Though see \citet{Wu_2014} which presented evidence of two distinct slopes in blue and red color regimes, i.e., two slopes with a break, for the HST F110W and F160W bands, which may be a genuinely anomalous effect for those bands \citep{Serenelli_2017}.}} However, at times, significant discrepancies in empirical zero points larger than 0.2~mag have been reported \citep{Valenti_2004, Gorski_2016}.}\footnote{{See \citet{Hoyt_2018} and \citet{Hoyt_2023} for resolutions to some of these discrepancies.}}

{In a joint study of IC~1613 and the LMC, \citet{Madore_2018} and \citet{Hoyt_2018} painted an optimistic picture for use of the TRGB in the NIR. They determined a TRGB color slope in IC~1613 and subsequently used it to construct a high-precision map of the apparent line-of-sight depth of the LMC. In doing so, they confirmed the LMC's known NE-SW tilt with a field-to-field residual distance dispersion of $\sim $0.05~mag.}
{A similarly positive outlook was presented in \citet{McQuinn_2019} who presented TRGB magnitudes that were synthesized by passing PARSEC isochrone predictions through anticipated JWST filter curves.}

{Both of these prior studies emphasized that the NIR TRGB can be a precision distance indicator as long as a robust fiducial calibration of its variation with color is determined; the variation with color is primarily a metallicity and secondarily an age effect. In this study, we will report the successful detection of the NIR TRGB feature with JWST and demonstrate that we can measure its color dependence. We then demonstrate how the use of such a color correction can significantly improve both the accuracy and precision of a TRGB distance.}

In Section~2, we describe the observation planning of the JWST program, the reduction of the NGC~4536 images, {and the sample selection procedure.} {In Section~3, we present the successful detection of all three distance indicators.} In Section~4, we estimate the TRGB's color dependence in the F115W band, use it to estimate the TRGB's color-corrected magnitude, {then test an expanded range of slope values on the data. We discuss the results in Section~5 and conclude in Section~6.}

\begin{figure*}[t!]
  \centering
  \includegraphics[width=0.84\textwidth]{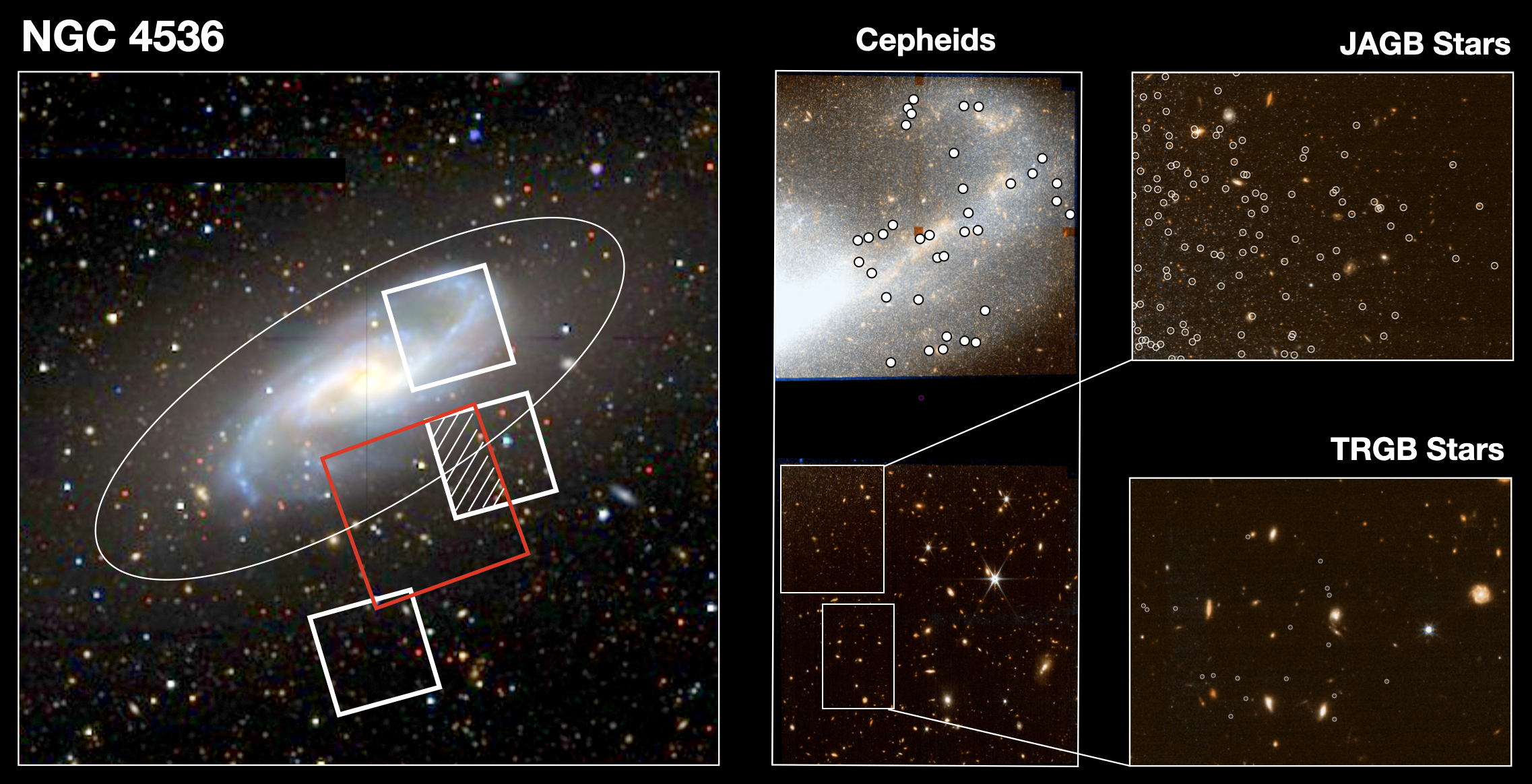}
  \caption{The successful imaging and detection of three stellar distance indicators in NGC~4536. \textit{left:} DECaLS \textit{gri} image. Footprints of the JWST/NIRCam (pair of white squares), NIRISS (singular white square), and archival HST/ACS (red square) imaging are plotted, with the overlap of the JWST and HST imaging marked (white hatched). North is up, East is left. The outer region of the galaxy adopted for TRGB analysis lies outside the indicated elliptical radius (white ellipse) of the galaxy's disk ($b/a=0.36$, $\theta = 120.7$~deg), which is adopted from HyperLEDA \citep[][\url{http://leda.univ-lyon1.fr/}]{Makarov_2014} and corresponds to a distance along its semi-major axis (SMA) equal to $7.5'$. \textit{middle column:} JWST/NIRCAM color image (R:F444W, G:(F115W+F444W)/2, B:F115W). NIRCam module A is aimed at the spiral arm where HII regions are apparent and the locations of Cepheid variables from \citet{Riess_2016, Riess_2022} are overplotted (white dots with black outline). In the other NIRCam module, plentiful, highly-reddened galaxies are seen in the background. Two regions are highlighted (white rectangles) to emphasize where each of the JAGB and TRGB stars are well-measured, respectively. \textit{Right:} Zoomed views of the two rectangular regions. The carbon-enhanced JAGB stars (top rectangle) appear red and are clearly distinguished from the RGB stars which appear blue (relative to $4.4~\mu\mathrm{m}$). The JAGB stellar density also falls off more quickly in radial distance than that of the RGB's due to the galaxy's outer disk profile being steeper than that of its stellar halo.
  }
  \label{fig:pointings}
\end{figure*}

\section{Observations and Data} \label{sect:data}

\subsection{Configuring the Program Observations}
As mentioned, JWST program GO-1995 aims to measure three different stellar distance indicators---Cepheids, TRGB, and JAGB---{simultaneously from one set of imaging per target.} Accomplishing this {imposed constraints on the placement of the scientific apertures, as well as on the allowed rotation angles of the telescope.}

For each target, the offset values of the NIRCam aperture were set so as to minimize loss of Cepheid coverage over large ranges of allowed telescope rotation angles, thereby maximizing schedulability at a minimal science loss. For observing the JAGB stars, additional constraints on the telescope's commanded position angles were determined using deep ground-based imaging from the DECALS legacy imaging survey \citep{Dey_2019}, as well as HI maps of the target galaxies when available. The goal was to sample each galaxy's thick disk component while minimizing the likelihood of dust extinction systematically biasing a JAGB measurement. Finally, (T)RGB stars were either targeted with {(some portion of)} the module of NIRCam that was not aimed at the Cepheids or with parallel NIRISS observations, depending on the angular extent of the target galaxy. In some cases, there existed ACS/WFC observations of the TRGB from the Carnegie Chicago Hubble Program \citep{Freedman_2019}. Sampling some portion of those imagesets was taken into consideration in order to leverage both optical and NIR photometry and better understand the TRGB magnitude-color relation (as in the present study).

JWST has complex, observatory-wide constraints that limit certain configuration parameters (e.g., absolute limits on the roll angle of the telescope given a target's position on the sky). In a small number of the program observations, this forced non-optimal configurations that, fortunately, did not significantly impact the science.
{Typical} examples include the drop out of one or a few Cepheid variables, or a reduced overlap with archival HST imaging. It is our recommendation that future investigators anticipate their exact roll angle requirements {as accurately and early as possible when planning a program so as to avoid having to make unanticipated trade-offs in the Phase 2 implementation.}

\subsection{Image Processing and Photometry}
We downloaded from MAST the 90 NIRCam integrations of our field in NGC~4536 (\autoref{fig:pointings}). The JWST imageset is composed of nine dithers mapped to each of the ten NIRCam detectors: eight in the short wavelength (SW) and two in the long wavelength (LW) channels. Each integration was comprised of six groups, resulting in an effective exposure time of 2802.297~sec in the F115W (SW) and F444W (LW) bands {acquired at one epoch of the imaged Cepheids' light curves}.

We processed the image data from \textit{raw} to calibrated \textit{cal} and undistorted \textit{i2d} images using the \textit{jwst} module version v1.8.2. {The JWST calibration  pipeline version number is implicitly associated with a set of calibration reference files such as the reference dark and flat images, amongst other image processing and calibration information.}
Two independent processing scripts, one from the command line and one in Jupyter Notebook, were implemented with default settings and confirmed to match exactly the MAST data products from the automated STScI processing pipeline.

The aligned F115W and F444W \textit{cal} images were then photometered with the DOLPHOT NIRCAM module \citep{Weisz_2023, Weisz_2024} that was developed as part of the JWST Early Release Science program on Stellar Populations (PI: Weisz). {At the time of the present analysis, the JWST DOLPHOT module was not fully released, and we used a beta early-release version that was graciously provided by the JWST-GO-1334 ERS team (PI: Weisz). The version used implicitly incorporated the critical updates to the NIRCam detector zero-points that were implemented in pmap0989 \citep{Boyer_2022}.} Aperture corrections were made using the DOLPHOT routine corresponding to ApCor=1. 

The archival CCHP HST/WFC data (GO-13691, PI: Freedman) acquired in the F606W and F814W bands were also reduced using DOLPHOT with a custom aperture correction routine applied that is expected to be more accurate than the automated DOLPHOT routines {in the case of deep HST photometry \citep{Jang_2023}.}

{Note that the JWST/NIRCAM photometry used in the analysis presented here was ``blinded'' by injecting into the photometry catalog a random, uniformly distributed offset between -0.2 and +0.2 mag, so the absolute flux zero points of NIRCAM do not play a role. Only the \textit{relative} intra-module, detector-to-detector, calibration would impact our analysis, and the post-pmap0989 detector-to-detector zero point uncertainties are at or below the measurement uncertainties discussed here. This is because any shifts to the absolute photometric zeropoints, the dominant uncertainty in the ongoing JWST flux calibration, amount to \textit{translations} on the color-magnitude diagram, thereby not distorting the TRGB morphology, i.e., its slope with color.}

Both HST and JWST catalogs were trimmed for well-measured point sources as described in \citet{Jang_2023}. The two catalogs were then cross-matched in WCS with a $0.12''$ radius threshold using methods from the \textit{astropy.coordinates} module. There was an approximate $0.05''$ offset in both RA and Dec between the default ACS and NIRCam WCS solutions, {with no evidence of higher order functionality in the aligned coordinate residuals}. The offset corrections were applied and the matching radius reduced to $0.08''$. The final dispersion about each matched set of coordinates was $0.01''$, or $<0.2$~pixels for each instrument.

The photometry was extinction corrected using a foreground reddening of  $E(V-I) = \eviGALONE \pm \eviGALONEerr$~mag adopted from the \citet{Schlafly_2011} recalibration of the \citet{Schlegel_1998} maps, as compiled in the NED extragalactic database (\url{ned.ipac.caltech.edu}). This corresponds to $A_{F606W}=\VbandextinctionGALONE$~mag, $A_{F814W}=\IbandextinctionGALONE$~mag, $A_{F115W}=\JbandextinctionGALONE$~mag, and $A_{F444W}=\FourFiveextinctionGALONE$~mag. An uncertainty equal to half of the reddening value is adopted due to the increased uncertainty associated with measuring MW dust in regions of low column density (i.e., at high galactic latitude).

\begin{figure}
    \centering
    \includegraphics[width=0.85\columnwidth]{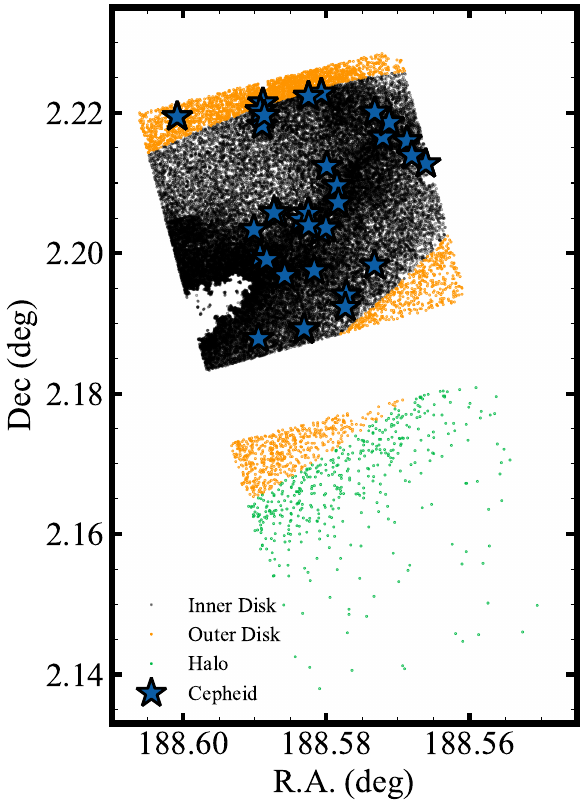}
    \caption{Sources brighter than the approximately identified TRGB magnitude (see \autoref{fig:cmd}) are plotted as dots. The spiral arm is clearly revealed in the upper module and contained in the inner disk selection (black dots). The dearth of sources in the outermost region of the imaging (green dots) is the hallmark of an old stellar population, i.e., a spatial manifestation of the TRGB. {The intermediate region is identified as the ``outer disk'' (orange dots). }{The interplay between this diagnostic and \autoref{fig:cmd} was used to converge on spatial selections for the JAGB and TRGB measurements}.
    }
    \label{fig:spatial_density}
\end{figure}

\subsection{Sample Definitions and Spatial Selections}

{Due to the different underlying physics, each distance indicator's population can be found in distinct regions on both the H-R diagram and in the spatial distribution of a galaxy's stellar mass. The Cepheids are found in the star-forming spiral arms and thin disk, the JAGB in the thermally-dispersed thick disk, and TRGB the diffuse stellar halo.} Therefore, we can {iteratively converge on an optimal spatial selection by examining the resultant CMD until each distance indicator appears most accurately measured.}

\begin{figure}
    \centering
    \includegraphics[width=0.95\columnwidth]{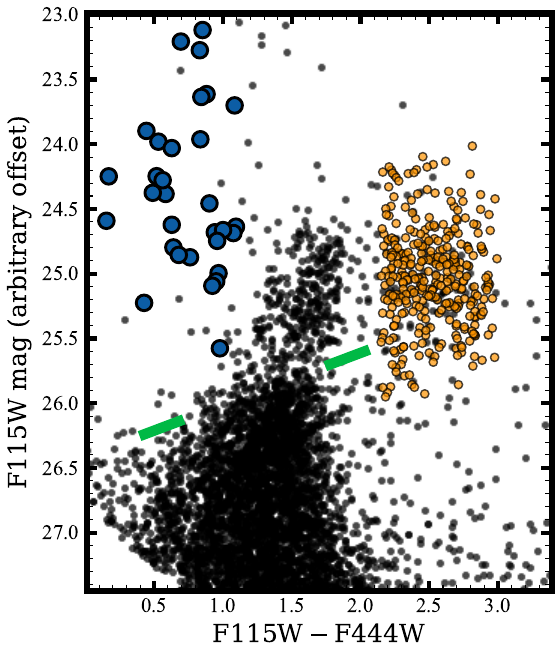}
    \caption{Simultaneous identification of three stellar distance indicators in NGC~4536. An overlay of three distinct stellar populations on the CMD. Black points are all stars contained in the halo selection (see \autoref{fig:spatial_density}), with the green lines outlining the TRGB feature (explored in detail later). Orange points are the JAGB stars selected in the outer disk region ($F115W-F444W > 2.2$~mag). Blue points are Cepheids, which predominantly occupy the inner disk region.}
    \label{fig:cmd}
\end{figure}

{This process is illustrated in \autoref{fig:spatial_density}, which plots the locations of all sources in the JWST catalog that are brighter than what is estimated to be the magnitude of the TRGB (green lines in \autoref{fig:cmd}.)} {The spiral arm stands out as a clear overdensity amidst the black points, consistent with it containing the youngest and brightest stars.} Outside of the spiral arm is a component that is slightly more diffuse and smooth in its structure, {but still has a measurable number of bright stars embedded in it, i.e., an intermediate-age population. Finally, at the largest separation from the disk components we start to see} {a constant and negligible density of stars brighter than the TRGB}, which is indicative of an old stellar population.

{These three regions delineate where each of the three standard candles is best measured. Though in the case of the young, massive Cepheids, the star-forming disk and spiral arm structures of galaxies are simply the \textit{only} regions in which they can be found.}

{The starting sample of Cepheids is the set union of those discovered and reported by two iterations of the SH0ES project \citep[][hereafter H16 and R22]{Hoffmann_2016, Riess_2022}. The periods were derived using WFPC2 observations acquired as part of the Sandage H0 project that was conducted at the same time as the HST-KP \citep{1994hst..prop.5427S, Saha_1996}. These were then followed up by the SH0ES program to place the Cepheid magnitudes onto the WFC3 system and update periods as needed \citep{Riess_2009}.}

\begin{figure}
    \centering
    \includegraphics[width=0.98\columnwidth]{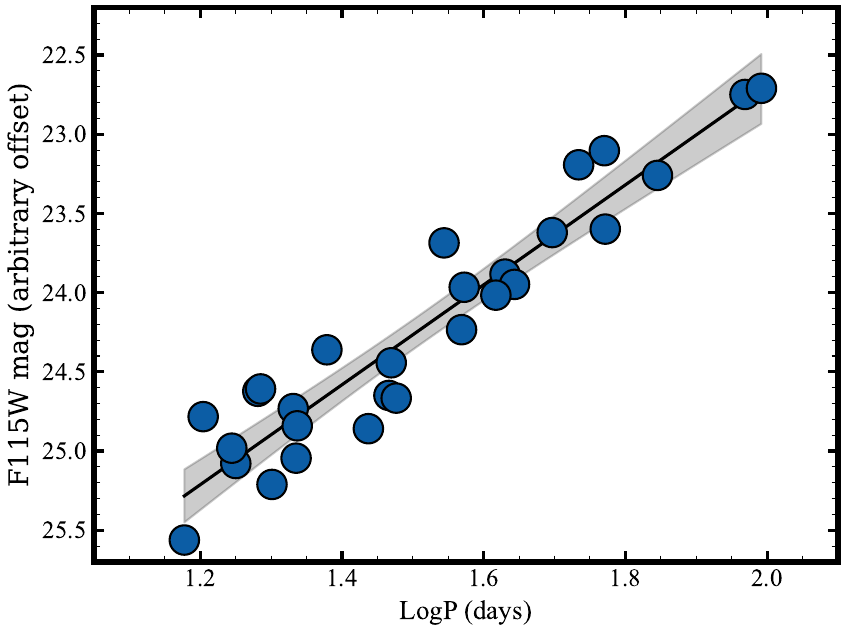}
    \caption{{Cepheid PL relation constructed directly from the measured F115W magnitudes and periods from the latest SH0ES sample of 28 Cepheids \citep{Riess_2022}. No correction has been applied for local scene crowding \citep[see, e.g.,][for a description]{Riess_2016} nor for extinction. The value of the best-fit slope is $-3.15 \pm 0.21$~mag/mag, and the dispersion about that slope is $0.240\pm0.024$~mag. The typical F115W photometric uncertainties are 0.015~mag and smaller than the plotted markers.}
    }
    \label{fig:ceph_plr}
\end{figure}

{We found significant shifts of $\delta \mathrm{RA} = -0\farcm59$ and $\delta \mathrm{Dec.} = 1\farcm24$ had to be added to the H16 coordinates to align them with the R22 ones. The R22 coordinates were then shifted by $\delta \mathrm{RA} = -0\farcs092$ and $\delta \mathrm{Dec.} = 0\farcs510$ to align them with our JWST WCS solution at a critical matching radius of $0\farcs08$. Doing so resulted in a sample of 35 Cepheids with periods ranging from 10d to 100d. All 35 cross-matched Cepheids were included in H16 and 28 in R22. This particular sample of Cepheids is very well measured relative to other more crowded datasets, so the possibility for false negatives or false positives in the cross-identification is very low. This can be verified via inspection of the Cepheid PLR (\autoref{fig:ceph_plr}). The Cepheid sample is already limited in numbers, and the number of Cepheids with well-measured F444W magnitudes was smaller yet (about 3-5), so we do not consider that band for the Cepheids here.}

{The other two indicators are best measured outside of the blue, star-forming regions of galaxies. See, e.g., \citet{Beaton_2019,Jang_2021,Hoyt_2021,Wu_2023} for TRGB and \citet{Lee_2024} for JAGB. Therefore, we elect to use the B-band surface brightness contours to parametrize our spatial selection, with the intent to mask the bluest star-forming regions. The adopted ellipse profile is the $D_{25}$ B-band isophote presented in HyperLEDA, with a $b/a=0.36$ and $\theta = 120.7$~deg \citep[][\url{http://leda.univ-lyon1.fr/}]{Makarov_2014}.}

{With this parametrization, we select for the JAGB an ``outer disk'' component at $SMA > 3\farcm2$ and for the TRGB an old/halo component at $SMA > 7\farcm5$. For the JAGB the SMA was varied until the population was clearly visible and distinct by-eye on the CMD (see orange dots in \autoref{fig:cmd}). JAGB stars were then further isolated on the CMD for colors $ 2.2 < (F115W-F444W) < 3.0$~mag and magnitudes $24.0 < F115W < 26.0$~mag. The spatial selection is plotted as orange dots in \autoref{fig:spatial_density} and the color-selected JAGB stars as orange circles in \autoref{fig:cmd}.}

{For the TRGB, the same was done but for the contrast of the TRGB as observed in the overlapping HST F814W-band imaging from which the TRGB has previously been identified \citep{Freedman_2019}.} As can be seen from the overlay with ground-based imaging in \autoref{fig:pointings}, the adopted halo-TRGB selection is consistent with a steep decline in diffuse light seen in the outer regions of NGC~4536, providing an unresolved consistency check on the resolved-star halo selection. {The adopted halo region is marked by green dots in \autoref{fig:spatial_density} and the approximately-identified TRGB bracketed by two green lines in \autoref{fig:cmd}.
}

\section{Identification and Detection of the Three Standard Candles}

\subsection{Cepheids}

{The Cepheid periods are zeroed to a fiducial $\log P = 1.50$ because the mean (log) period of the H16 sample is 1.48 and the mean of the R22 sample is 1.51. The slopes inferred from the H16 ($N=35$) and R22 ($N=28$) Cepheid samples are, respectively, $-3.01 \pm 0.19$~mag/mag and  $-3.15 \pm 0.21$~mag/mag. The intercepts are $24.239 \pm 0.043$~mag and $24.266 \pm 0.047$~mag. The dispersions about the best-fit lines are $0.246 \pm 0.020$~mag and $0.240 \pm 0.024$~mag. The 28 Cepheids in R22 are a perfect subset of the 35 in the H16 sample, so the results are highly covariant.}

{Note our analysis here does not include vital corrections for crowding and internal extinction (only foreground) because the evidence for both effects is near zero for this galaxy's Cepheids (one of the most nearby and least crowded in the sample). A more complete treatment of the Cepheid PLR in these bands for the more distant Cepheids, and which supersedes this preliminary analysis, can be found in Owens et al. (2024, in review).}

{The Leavitt slopes derived from either sample of Cepheids in NGC~4536 (SH0ES 2016 or 2022) are consistent with the value of $3.15 \pm 0.07$~mag/mag observed in the UKIRT/2MASS $J$-band (separated in effective wavelength from F115W by only 90~nm). Furthermore, the measured dispersion is consistent with previous findings for random-phase sampling of Cepheids in the similar $J$-band \citep{Persson_2004, Monson_2012}.
}

\subsection{JAGB/Carbon stars}

\begin{figure}
    \centering
    \includegraphics[width=0.95\columnwidth]{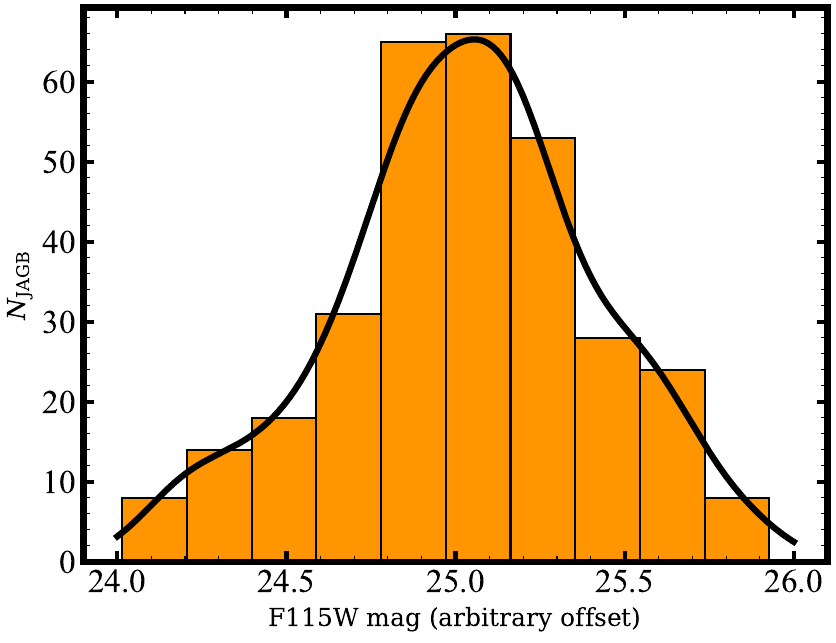}
    \caption{{JAGB Carbon Star F115W luminosity function, i.e., marginalizing over color for the orange points in \autoref{fig:cmd}. The black curve is a KDE estimate of the distribution with modal value equal to $25.036 \pm 0.056$~mag.}}
    \label{fig:jagb_lf}
\end{figure}

{In \autoref{fig:jagb_lf}, the JAGB luminosity function (LF) that was selected for via color and magnitude (orange disks in \autoref{fig:cmd}) is shown. The standard deviation of the distribution is $0.381 \pm 0.014$~mag and the normalized median absolute deviation (NMAD) is $0.340 \pm 0.025$~mag. Being based on median statistics, the NMAD is more resilient to outliers and asymmetric or non-Gaussian structure in the tails of a distribution. Typical JAGB LF widths are between 0.30~mag and 0.36~mag \citep[see, e.g.,][]{Lee_2024_fourstar}, confirming that we have identified the JAGB LF.}

{To determine the JAGB magnitude, we first compute a kernel density estimate (KDE) of the JAGB LF. We then resample the JAGB LF and a new modal magnitude is computed until convergence is reached in the median of resampled modal values (found to occur at $N_{resamp} \geq 300$). The median and standard deviation of resampled modes are taken as the central value and uncertainty, respectively, resulting in an $\mathrm{F115W}_{JAGB} = 25.036 \pm 0.056$~mag.}

{Our treatment of the JAGB here is superseded by the dedicated analysis from \citet{Lee_2024}, which includes tests of the spatial and color selection. We do not undertake those here, and instead have simply adopted values that produce an expected JAGB population on the CMD. The aim here is not to explore the method's uncertainties at the percent level, but to demonstrate its simplicity, that the population of stars in question have been detected in the JWST imaging, and that a precision distance can be determined.}

\begin{figure*}
    \centering
    \includegraphics[width=0.95\textwidth]{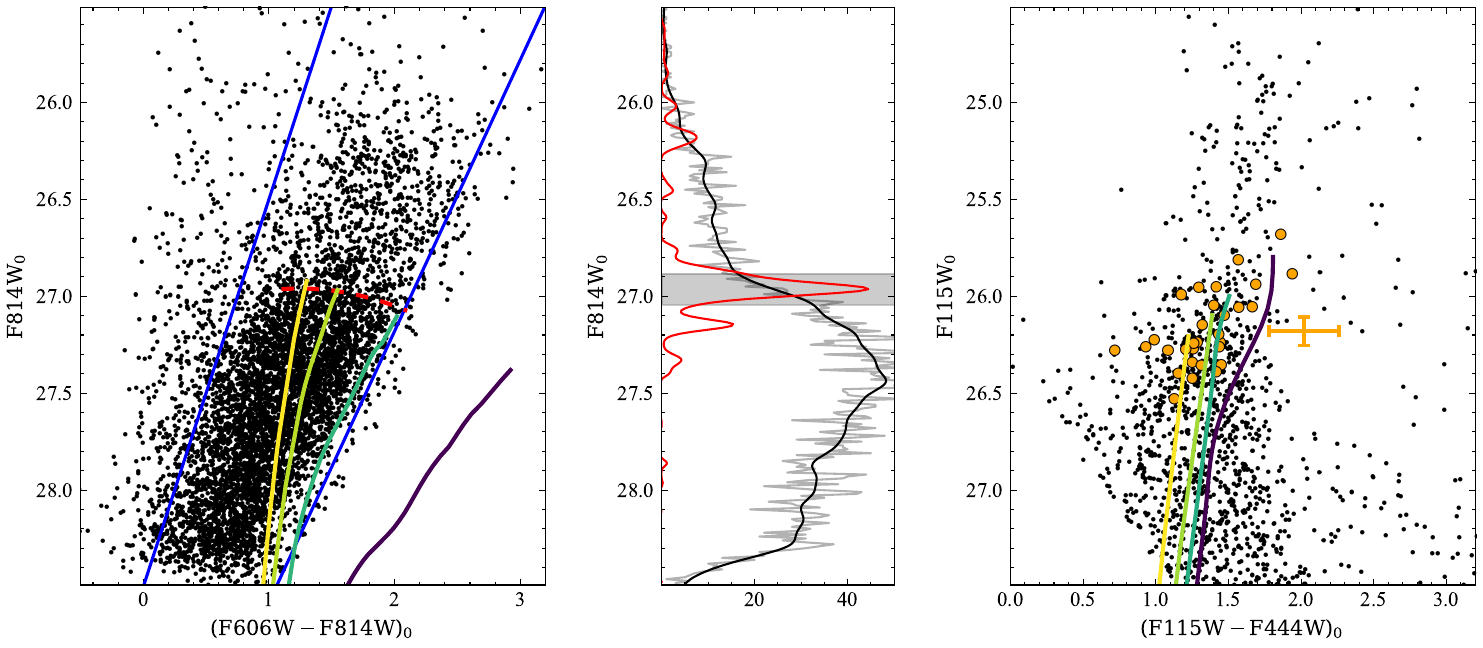}
    \caption{Identification of TRGB stars in NGC~4536 from the overlap of the optical ACS/WFC and infrared NIRCam fields of view. \textit{left}: Optical CMD. The quadratic F814W-band color dependence from \citet{Jang_2017_color} was adopted {(red dashed curve)} to rectify, or flatten, against color the TRGB feature contained within the selection of RGB stars (blue slanting lines). Overplotted are 10~Gyr BASTI isochrones arbitrarily scaled along the vertical axis and colored (light yellow to dark purple) according to their metallicities $Z = \{0.002, 0.004, 0.008, Z_{\odot}\}$.
    \textit{Middle:} The edge detector response (EDR, red curve) function is is computed as the Poisson-weighted first derivative of the rectified luminosity function (LF, gray and black curve). Tip stars are identified within the $\pm \text{\bf0.08}$~mag window centered on the location of the TRGB peak in the EDR.
    \textit{Right}: The F814W-identified TRGB stars are highlighted in the JWST/NIRCAM CMD. Overplotted are again 10~Gyr BaSTI isochrones, this time as predicted for the JWST NIR filters in this study. {A set of representative error bars for the TRGB stars is plotted (orange cross with caps).} A single random offset between $-0.2$ and $0.2$~mag has been applied to both of the F115W and F444W magnitudes.
    }
    \label{fig:tip_id}
\end{figure*}

\subsection{Identifying TRGB Stars}

To identify TRGB stars in the IR CMD we look to the F814W-band ($\lambda_{eff} \simeq 800$~nm) where the TRGB magnitude's dependence on photometric color is significantly reduced relative to other wavelengths. The F814W magnitudes are rotated to a coordinate grid on which the TRGB's (small) color dependence is ostensibly flattened \citep[or ``rectification,'' as introduced in][]{Madore_2009}. The specific transformation adopted here is the $QT$ (quadratic TRGB) equation of \citet{Jang_2017_color}, which has been shown to be most consistent with model predictions \citep[e.g.,][]{Serenelli_2017} and with high-precision measurements of the TRGB made in the Magellanic Clouds \citep{Hoyt_2023}. A TRGB magnitude is then measured from the $QT$ magnitude luminosity function (LF) according to the methodology described in \citet{Hoyt_2023} and also used by the Carnegie Chicago Hubble Program \citep{Freedman_2019}.

In \autoref{fig:tip_id}, the identification of TRGB stars is illustrated. First, as shown in the left-panel of \autoref{fig:tip_id}, RGB stars are selected from the F814W, F606W CMD via a blue and red color cut plotted as blue lines. The magnitudes of these stars are then rotated to flattened $QT$(814) magnitudes and used to construct a new RGB LF from which the TRGB discontinuity is to be detected via computation of the Poisson-weighted first derivative, or EDR. The width of the TRGB peak computed from the rectified RGB LF was 15\% narrower than that measured from in the unrectified RGB LF, thereby justifying the use of a color correction. For reference, BaSTI isochrones with age 10~Gyr are shown and color-coded for Z = 0.002, 0.004, 0.008, and Solar metallicity \citep{Hidalgo_2018, Pietrinferni_2021}. {The curvature of the isochrone-predicted F814W TRGB agrees well with the empirical JL17 QT calibration. This underscores that a second-order color correction is likely the best to use when making precision, color-corrected TRGB distances in the F814W band.}

\begin{figure*}
    \centering
    \includegraphics[width=0.95\textwidth]{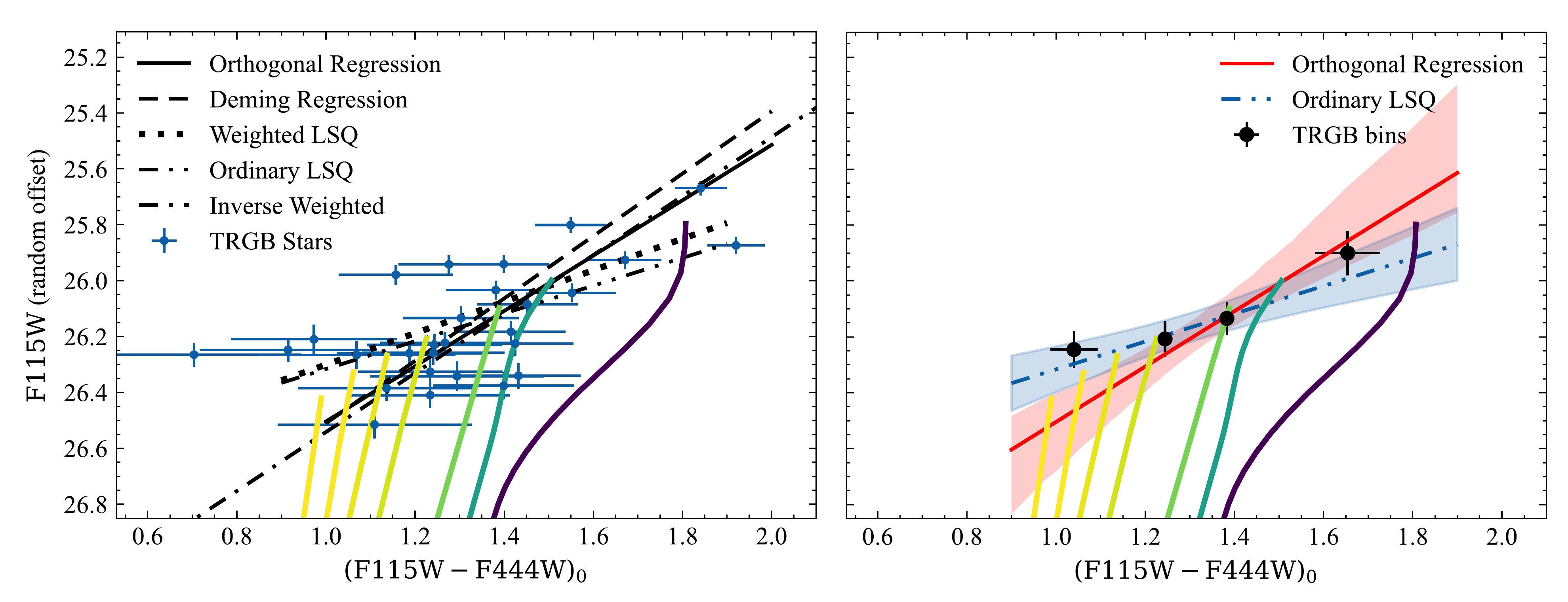}
    \caption{{Fits to the TRGB's magnitude-color relation in the F115W and F444W bandpasses}. \textit{left:} F115W magnitudes versus $(\mathrm{F115W} - \mathrm{F444W})$ colors for tip stars (blue circles) previously shown as orange circles in \autoref{fig:tip_id}, now plotted with their photometric errors (blue lines). Note that most of the scatter about the trendlines is due to color uncertainties from the F444W magnitudes. \textit{Right:} {Same as left panel, with four equal-numbers bins (sorted by color) representing the TRGB magnitudes and colors (black dots). The errorbars represent the standard deviation of each quantity's distribution within each bin. Two of the fitted lines are plotted along with their 90\% CI bands determined via bootstrapping: the result from ordinary least squares (blue dot-dot-dashed line amd band) and orthogonal distance regression (red solid line and band). The fits were performed on the unbinned datapoints (left), not the binned points (right).}
    In both panels, BaSTI 10~Gyr isochrones are again plotted on a Viridis color map, this time over a metallicity range expanded to include more metal-poor isochrones, $Z = \{0.0002, 0.0004, 0.0008, 0.0014, 0.0039, 0.008, 0.017\}$ or $[\mathrm{Fe}/\mathrm{H}] = \{-1.90, -1.55, -1.30, -1.05, -0.60, -0.30, +0.06\}$ 
    }
    
    \label{fig:slope_errs}
\end{figure*}

In the middle panel of \autoref{fig:tip_id}, the $QT$-mag EDR is plotted as a red curve along with the F814W smoothed and unsmoothed RGB LFs as black and gray curves, respectively. A transparent, gray, horizontal band represents the window that was used to select TRGB stars within {$\pm 0.08$~mag of the peak's location, set by the typical F814W magnitude uncertainty at the magnitude of the TRGB. Smaller and larger windows than this were tested and it was found that windows smaller than about 0.04~mag suffered from too small number statistics, and windows larger than 0.08~mag caused the TRGB slope fit to become increasingly} skewed from the 2-dimensional edge feature towards tracing the steep slope of the RGB itself, {i.e., contamination from non-tip stars became too severe. So the 0.08~mag window was the right balance of number statistics and sample purity and also motivated by the actual measurement uncertainties. Note that all of the below analysis was also performed using a 0.05~mag window, which is the width of the F814W TRGB peak. The findings were consistent with the presented ones for a 0.08 mag window but based on exactly half the numbers of tip stars.}

In the right panel of \autoref{fig:tip_id}, the NIRCAM JWST CMD is shown, with the tip stars marked as orange. Also plotted are the same 10~Gyr BaSTI isochrones but for the JWST/NIRCAM filters. These stars appear to form a sloped sequence along the upper edge of the RGB, which we conclude to be the TRGB. In the following section, will use these traced stars to derive a value for the color dependence of the JWST TRGB. 

{Note the MCR-TRGB methodology introduced in \citet{Durbin_2020} provides a more rigorous approach to simultaneous determination of the multi-band TRGB relation than is done here. However, its advantages can only be realized when applied to datasets with much higher photometric signal-to-noise at the TRGB than is viable {for HST to reach} at the distances of the SN Host galaxies like that presented in this study. It will be prudent to deploy the MCR-TRGB methodology once a sufficiently large sample of TRGB stars observed with both HST and JWST at high SNR is attained \citep[e.g.,][]{2021hst..prop16743H, 2021jwst.prop.1638M}.}

\section{TRGB Measurement}

{In this section, we use the TRGB stars that were selected in the HST/ACS/F814W band (see previous section) to estimate the slope of the IR TRGB's magnitude-color relation as observed in the NIRCAM bands. We then expand the sample to the full JWST footprint of NGC~4536 and estimate the TRGB via rectification (rotation) of the NIRCAM CMD using the newly-determined slope. Finally, we demonstrate the impact that different adopted slope values have on TRGB measurement in terms of the morphology of the EDR, the width of the dominant discontinuity feature, and the estimated distance modulus.}

\begin{deluxetable*}{lrrrrr}
\tabletypesize{\normalsize} 
\tablewidth{0pt} 
\tablecaption{TRGB Slope Determinations\label{tab:slopes}} 
\tablehead{ 
\colhead{Fits to JWST TRGB Data} &
\colhead{Slope} &
\colhead{$\sigma_{slope}$} &
\colhead{$\Delta y_0$\tablenotemark{a}} &
\colhead{$\sigma_{y_0}$} &
\colhead{$x_0$} 
}
\startdata 
Ordinary Least Squares                 & $\ordinaryLSQval$ & \ordinaryLSQerr &  0.01 &  0.03 &  1.29 \\
Weighted Least Squares                   & $\weightedLSQval$ & \weightedLSQerr & -0.03 &  0.03 &  1.29 \\
Inverted Least Squares\tablenotemark{b}  & $\invertedLSQval$ & \invertedLSQerr & -0.01 &  0.04 &  1.36 \\
Deming Regression\tablenotemark{a}       & $\Demingval$ & \Demingerr & -0.01 &  0.04 &  1.32 \\
\textbf{Orthogonal Distance Regression}           & $\ODRval$ & \ODRerr &  0.03 &  0.05 &  1.29 \\
\hline\hline
BaSTI-predicted Metallicity Slopes & [Fe/H] & $= [-1.90,$ & $+0.06]$ &dex \\
\hline
$\mathrm{Age}=10$~Gyr  & $-0.74$ & 0.03 & \nodata & \nodata & \nodata \\
$\mathrm{Age}=~4$~Gyr  & $-0.92$ & 0.07 & \nodata & \nodata & \nodata 
\enddata 
\tablecomments{\textbf{The preferred estimate from Orthogonal Distance Regression (ODR) is bolded.}}
\tablenotetext{a}{Intercepts zeroed to the mean across all fits and reported as shifts from that mean value.}
\tablenotetext{b}{The slope reported here was computed as the inverse of the best-fit value inferred via regression.}
\tablenotetext{c}{68\% confidence interval determined via bootstrapping. CI was symmetric so quoted here as scalar-valued.}
\end{deluxetable*}

\subsection{Determining the TRGB Magnitude-color Relation}
With TRGB stars identified from the F814W$_0$ vs. (F606W$-$F814W)$_0$ CMD, they can be immediately traced to those stars for which counterparts were found in the JWST F115W and F444W photometry. We can then infer the parameters of a linear relationship between the F115W magnitudes and (F115W$-$F444W) colors of the TRGB stars. We present multiple approaches to fitting a line to data (from ordinary least squares to orthogonal distance regression), which differ in their treatment of uncertainties.
In all cases, the colors of the tip stars are zeroed to their mean (F115W$-$F444W) color (1.31~mag).

\subsubsection{Least Squares}
{Three forms of least squares fits are performed: ordinary, weighted, and inverse-weighted, resulting in slopes $\ordinaryLSQval \pm {\ordinaryLSQerr}$~mag/mag, 
${\weightedLSQval} \pm {\weightedLSQerr}$~mag/mag, and
${\invertedLSQval} \pm {\invertedLSQerr}$~mag/mag. The inverted least squares was a common approach in Astronomy that was meant to estimate the bias that large uncertainties on the independent variable axis can introduce into weighted fits, done so by simply flipping the dependent and independent axes in a weighted least squares fit. The fit results from each are printed as the first three rows of \autoref{tab:slopes} and plotted in the left panel of \autoref{fig:slope_errs} as the dash-dot-dot, dotted, and dash-dot-dash lines for ordinary, weighted, and inverse-weighted.}

\subsubsection{Deming Regression}
In the 2-D case, when measurement uncertainties along both axes of variation are nonzero and disproportionately-sized, any perceived correlation will become elongated along the axis with larger measurement uncertainties.\footnote{\citet{Spearman_1904} referred to this as ``constriction'' and/or ``regression dilution.'' The effect was independently re-discovered by \citet{Brodie_1980} in regards to the Cepheid period-luminosity relation.} An attempt to correct for this ``Diffusion Bias'' is the Deming regression \citep{Deming_1964}, which uses prior information on the ratio of expected variances due to measurement errors $\sigma_x / \sigma_y$ to mitigate biases introduced when regressing against data with non-zero measurement errors along both axes of variation. 

From the Deming regression, and an assumed variance ratio of 9 we find a slope equal to ${\Demingval} \pm {\Demingerr}$~mag/mag. The uncertainty intervals are estimated via bootstrapping for 10000 iterations. This inference should provide a better comparison with theory, or with measurements of the TRGB slope made in the same bands but at higher photometric signal-to-noise, and which will more closely approach the ``true'' slope as the Deming regression aims to infer. The result is plotted as the dashed line in \autoref{fig:slope_errs} and is the fourth entry in \autoref{tab:slopes}.

\subsubsection{Orthogonal Distance Regression}
Finally, we undertake a more appropriate treatment of two-dimensional uncertainties via orthogonal distance regression (ODR). ODR minimizes the orthogonal distance of each point from the best-fit line, as opposed to just the vertical distance as in ordinary regression \citep[see, e.g.,][regarding its implementation in the Fortran package ODRPACK]{Boggs_1990}. The resulting best-fit slope is ${\ODRval} \pm {\ODRerr}$~mag/mag. This result is plotted as a solid line in \autoref{fig:slope_errs} and is the fifth entry in \autoref{tab:slopes}.

\begin{figure*}
    \centering
    \includegraphics[width=0.7\textwidth]{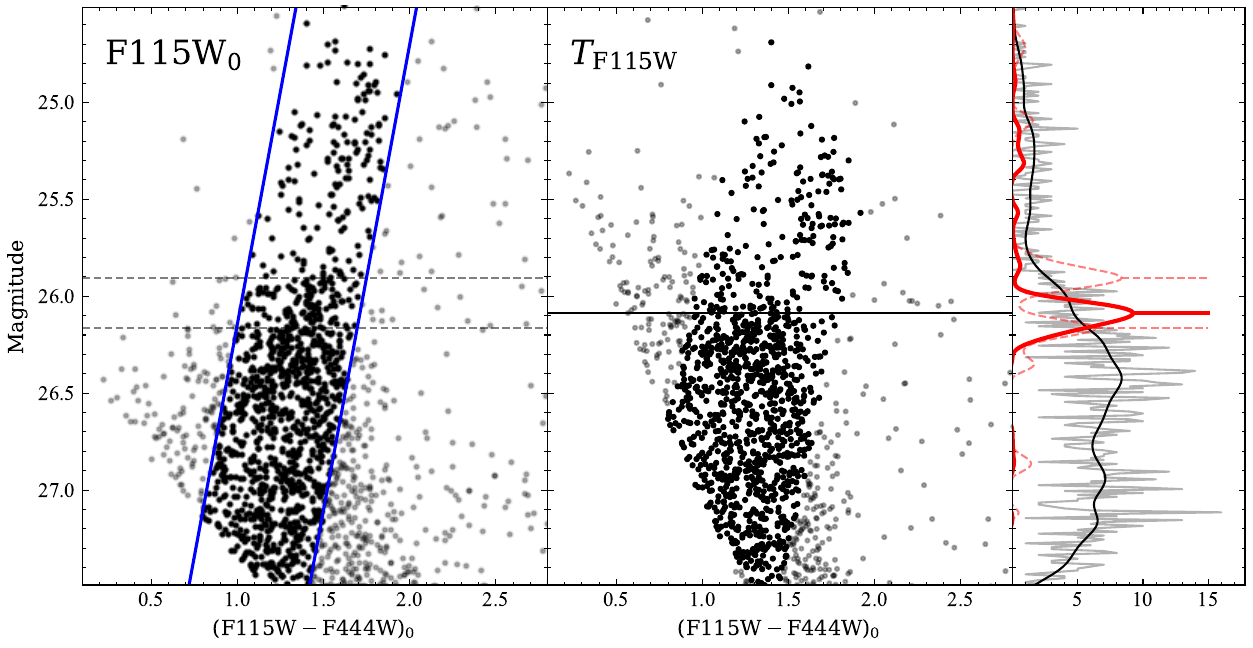}
    \caption{Demonstration of TRGB rectification in the full JWST halo dataset and subsequent sharpened measurement of the IR TRGB in JWST filters. \textit{Left}: F115W vs. F444W CMD. The color-magnitude selection used to select along the RGB is shown (blue lines) with RGB sources (solid black points) plotted along with non-RGB sources (transparent black points). The locations of two equally likely peaks in the resulting EDR (red dashed curves in right-most panel) are marked (transparent, horizontal dashed lines). \textit{Middle}: $T_{115}$ vs. F444W CMD, wherein the stars that fall inside the unrectified RGB selection (solid black if RGB, transparent gray if not) are rotated via separate derivation of the TRGB's color slope (adopted here as \ODRval~mag/mag), such as that determined in the previous section. The resulting EDR is singly-peaked, demonstrating the success of the TRGB slope in improving the clarity of the TRGB. The location of the peak magnitude is marked on the rectified CMD (horizontal black line). \textit{Right}: Unrectified RGB LF are computed for bins of size 0.01~mag (gray curve) and smoothed with Gaussian kernel of width 0.08~mag (black curve). Plotted also are the EDRs derived from both the unrectified (red, transparent, dashed curve) and the rectified (solid red curve) RGB magnitudes. Both EDRs have been normalized to the same peak value and scaled to the maximum value of the RGB LF. The locations of the two strong peaks in the unrectified EDR are marked (red, horizontal, dashed lines) along with the single dominant peak in the rectified EDR (red, horizontal, solid line).
    }
    \label{fig:rect_trgb_measure}
\end{figure*}

\subsubsection{BaSTI Isochrones}

{The BASTI isochrone suite was queried and used to estimate predicted slopes in the JWST bands considered here. The isochrone metallicities are:
 $[\mathrm{Fe}/\mathrm{H}] = \{-1.90, -1.55, -1.30, -1.05, -0.60, -0.30, +0.06\}$. For two set ages, 4~Gyr and 10~Gyr, the metallicity slope is then computed by fitting a line to the single brightest step at the Tip of each RGB isochrone. The 10~Gyr slope is estimated to be $-0.74 \pm 0.03$~mag/mag and the 4~Gyr slope $-0.92 \pm 0.07$~mag/mag. The uncertainties are the formal standard errors returned from the fit. Their being non-zero (despite being derived from an ``error-less'' prediction) is a function of the discrete sampling in metallicity space, and any deviations along that discrete sampling from a scatter-less line. In \autoref{fig:slope_errs}, the 10~Gyr isochrones are plotted on a viridis color-mapping. In \autoref{tab:slopes}, both slopes are presented in the last two rows. In Appendix A, the BaSTI-predicted TRGB metallicity slope is plotted from 2~Gyr to 13~Gyr in 1~Gyr steps, computed in the same way as just described.
 }
 
\subsection{Measuring the TRGB in NGC~4536}
In the previous sections, the region of overlap between the archival ACS/WFC and new NIRCam imaging (hatched region outside the ellipse in \autoref{fig:pointings}) was used to {estimate} the TRGB's color dependence. In this section, we can now incorporate the remainder of the NIRCam imaging and use the newly determined TRGB slope to ``rectify'' (or rotate such that the TRGB is flat in color) the CMD and measure the TRGB along the one-dimensional magnitude axis. {Simply, the F115W magnitudes are transformed as,}

\begin{equation}
    T_{115} = m_{115} - \beta (c - c_0)
\end{equation}
{where $\beta$ is the adopted TRGB slope, $c$ is a star's color index, and $c_0$ is the pivot color about which the RGB is being rotated.}

The area of the remaining NIRCAM data is about 1.5$\times$ larger than the subset which overlaps with HST, providing a good dataset to ``test'' the TRGB slope that was inferred from the HST+JWST ``training'' data. In other words, this section {demonstrates how one could use an externally-calibrated TRGB slope to derive a TRGB distance in these JWST bands.}

In \autoref{fig:rect_trgb_measure}, a TRGB measurement is demonstrated for a rectification plus 1D edge detection approach. In the left panel, the F115W vs. (F115W$-$F444W) CMD is shown, along with the RGB color-magnitude selection, and two horizontal lines depicting the two equally likely peaks seen in the unrectified EDR (right panel dashed transparent red lines). The F115W magnitudes are then rectified to $T_{115}$ magnitudes with slope $\beta = {\ODRval}$~mag/mag and a new RGB LF is constructed by marginalizing over color. The width of the TRGB peak in the EDR derived from the rectified LF is smaller than the separation of the two equal-strength peaks in the unrectified EDR, demonstrating the significant improvement in precision and accuracy when applying the newly-determined TRGB slope to the remainder of the JWST imaging of the NGC~4536 halo.

\subsection{Slope Variation Experiments}

\begin{figure*}
    \centering
    \includegraphics[width=0.75\textwidth]{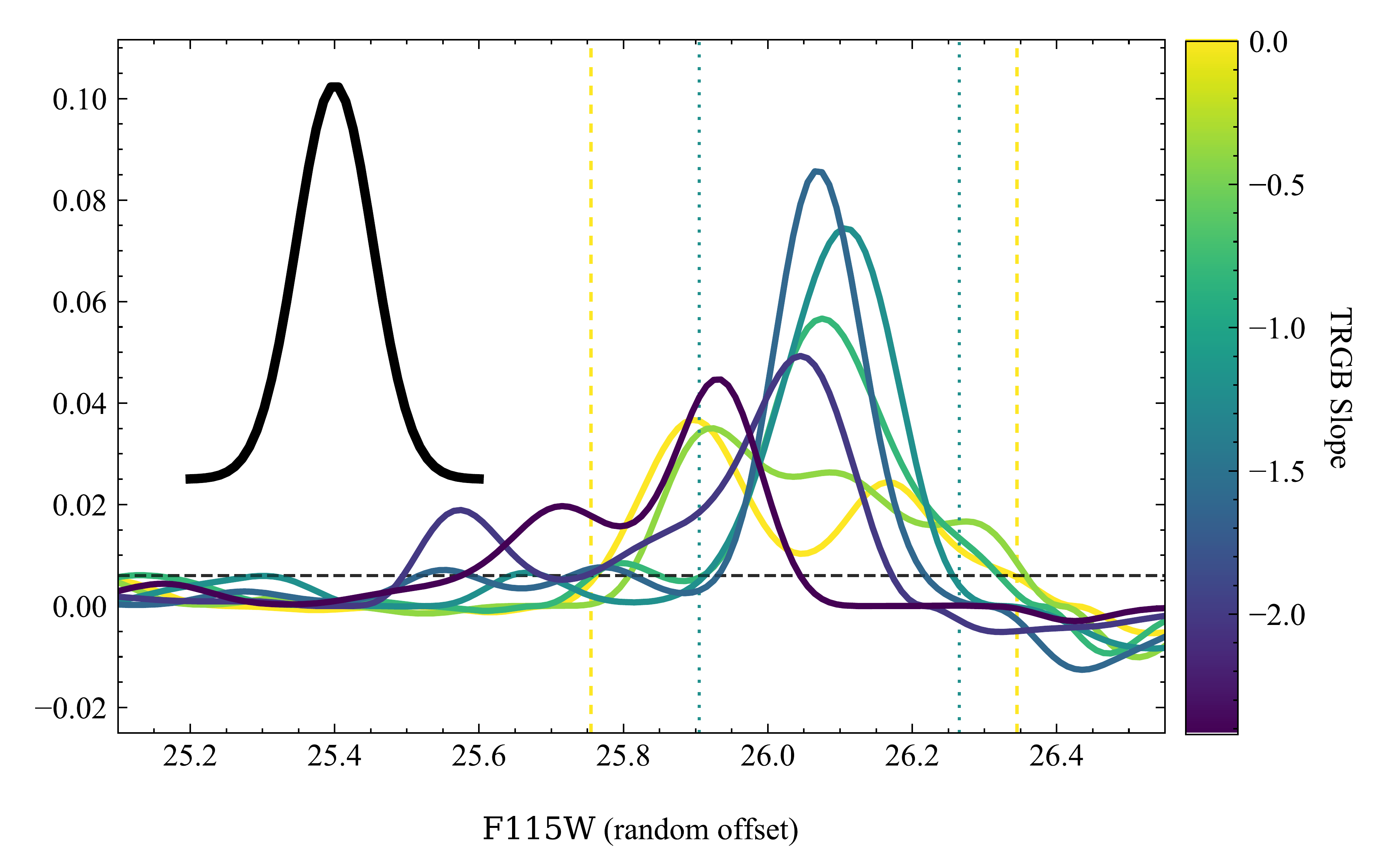}
    \caption{Edge Detector Response curves (EDRs) computed from RGB luminosity functions (RGB LFs) that were derived from rectified $T$ magnitudes over a range of TRGB slopes used for the rectification. {Seven representative EDRs are shown to demonstrate the behavior of vertical edge features in the rotated and color-marginalized CMD as a function of the slope used to perform said rotation. Plotted are the left and right edges of the uncorrected (slope equals zero) TRGB peak as well as those belonging to the peak resulting from the ODR slope determined in the previous section (vertical dashed lines, colored to match the associated EDR curve and slope value)}. The threshold adopted for defining the edges of peak features is also plotted (horizontal black dashed line). For reference, the kernel used to smooth the RGB LF is shown with arbitrary height (black curve, top right).}
    \label{fig:running_slope_edrs}
\end{figure*}

\begin{figure}
    \centering
    \includegraphics[width=0.98\columnwidth]{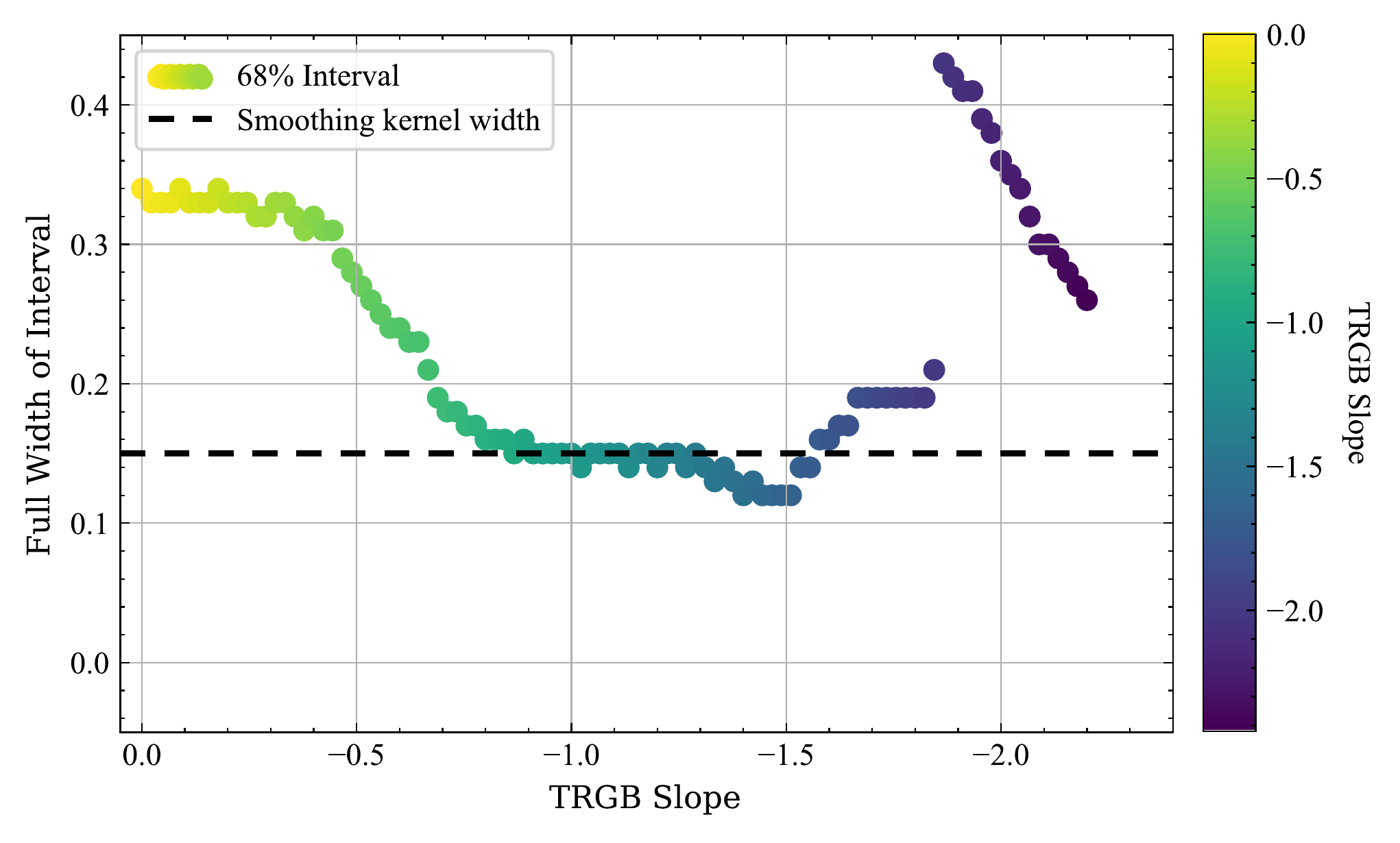}
    \caption{68\% width of the TRGB peak feature as a function of the slope that was used to flatten the TRGB feature on the CMD (rectification). {One hundred linearly spaced steps in TRGB slope values ranging from $-2.2$ to 0.0 were considered}. The width of the smoothing kernel is plotted for reference (horizontal black dashed line). {Note both the x-axis and the color map are aligned to the TRGB slope axis. The redundancy is meant to aid in comparison with the representative EDRs plotted in \autoref{fig:running_slope_edrs}}.
    }
    \label{fig:running_widths}
\end{figure}

\begin{figure}
    \centering
    \includegraphics[width=0.98\columnwidth]{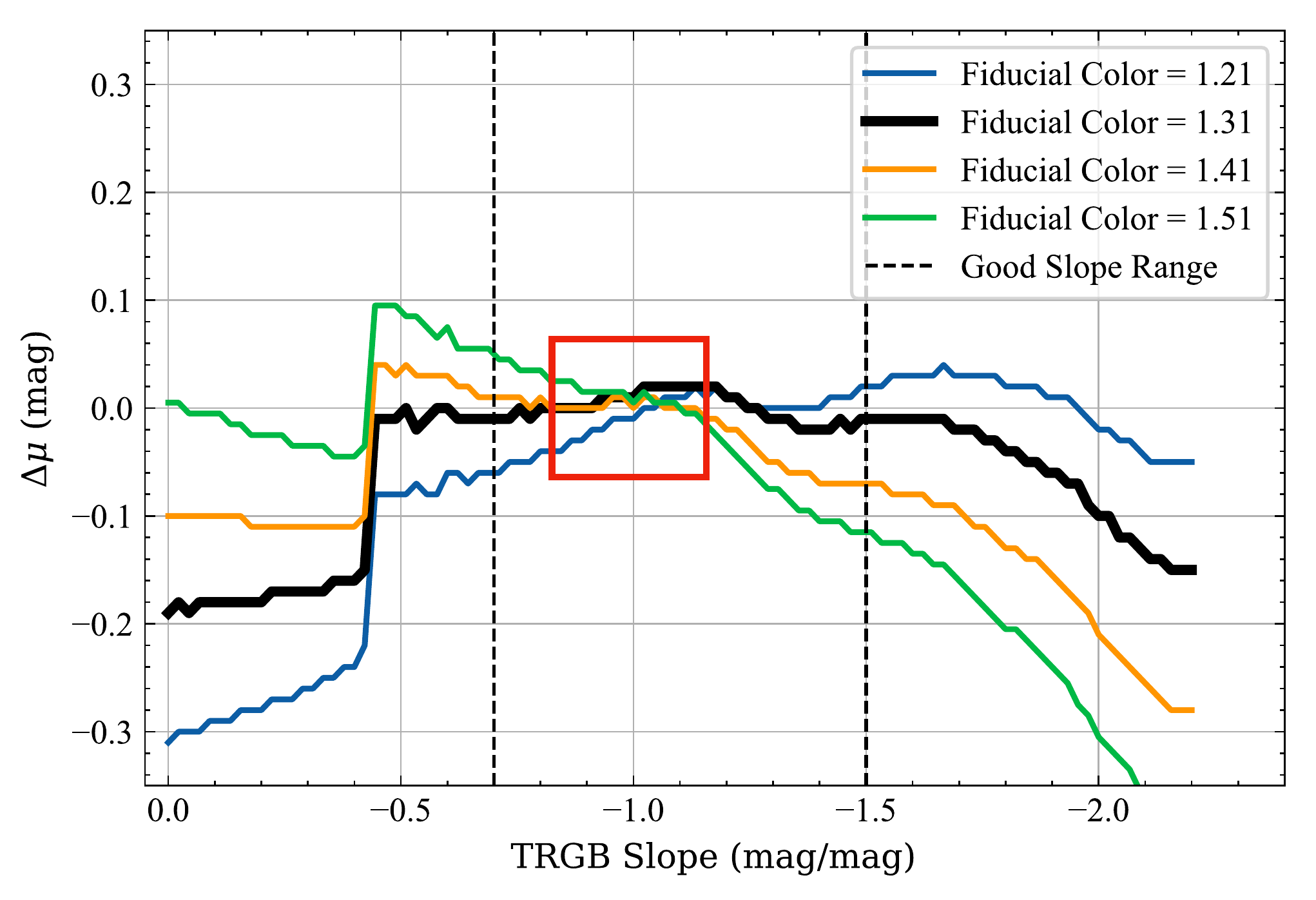}
    \caption{{Delta in TRGB distance as a function of adopted TRGB slope for four fiducial calibrations. The thick, black curve represents rotation of the CMD about the mean color of the observed TRGB. The thinner, colored curves represent different pivot colors. The distance shifts $\Delta \mu$ are computed relative to the median of the black curve within the ``good slope range'' (vertical lines), determined from the top panel. The red, boxed region encloses the suite of curves contained by the standard error interval of the ODR results from the optical-IR trace method.}
    }
    \label{fig:running_delta_mu_fidu}
\end{figure}

We further explore the TRGB slope with a more brute force approach to minimization. Over a grid of possible TRGB rectification slopes (from 0 to $-2.20$~mag/mag), we recompute the EDR and document the width of the dominant edge feature (sometimes observed as a cluster of multiple, equal-power peaks). {The results of this experiment are illustrated in \autoref{fig:running_slope_edrs} and summarized in Figures \ref{fig:running_widths} and \ref{fig:running_delta_mu_fidu}.}

In \autoref{fig:running_slope_edrs}, the EDRs as derived from {seven} representative sets of rectified F115W magnitudes are plotted and mapped to their unique slope values. The EDRs shown correspond to $\beta = \{-0.00, -0.37, -0.73, -1.10, -1.47, -1.83, -2.20\}$ mag/mag. {The exact color mapping is shown in the right-hand color bar.}
The significant sub-structure in the EDR for very shallow and large values of the TRGB slope, {are a result of under- or over- rotation of the CMD, signaling a slope incompatible with the data.}
For values of the TRGB slope between about -0.7 and -1.5, that substructure can be seen to merge into one sharp peak that approaches the minimum scale length of any feature in the EDR {that can be considered real and not a noise fluctuation}. This {floor in the feature} scale length is set by the size of the smoothing kernel {which is plotted for reference as a thick black curve in the top-left of \autoref{fig:running_slope_edrs}.}
The bounds within which the peak widths were computed are also shown for two representative EDRs, one {unrectified} and one rectified by the best-fit value --0.99 mag/mag estimated in the previous section. {The significantly reduced width of the EDR feature is a result} of the {merger of multiple peaks observed in the unrectified EDR} into a single dominant peak.

In \autoref{fig:running_widths}, this sharpening of the tip feature via rectification is shown by plotting the 68\% confidence interval widths of the TRGB peak feature as a function of adopted slope for rectification. The results show {that for slope values from about $-0.7$ to $-1.5$~mag/mag}, the observed dispersion in the TRGB peak approaches the minimum feature size set by the size of the smoothing kernel. The results agree well with the range of slopes presented in \autoref{tab:slopes}{, both the empirical estimates and BaSTI isochrone predictions.}

{In \autoref{fig:running_delta_mu_fidu}, the shift in estimated TRGB distance is plotted as a function of adopted TRGB slope for four hypothetical fiducial calibrations of the TRGB. That is, $\beta$ is varied continously and $c_0$ is set to four fiducial values. This is meant to approximate how the TRGB distance estimated to this galaxy might change depending on both the TRGB slope and the mean color of one's TRGB calibration dataset, e.g., the color of the TRGB in geometric anchors like NGC~4258 or the LMC. The shifts in distance $\Delta \mu$ are computed relative to the median of the TRGB magnitudes within the $-0.7$ to $-1.5$~mag/mag range of ``good'' slope values (see top panel and previous paragraph). The red boxed region encloses the suite of curves that lie within the standard error of the nominal ODR results for the TRGB slope in the previous section ($\beta = 0.99 \pm 0.16$ mag/mag). This provides an estimate of the additional contribution to the TRGB distance error incurred by extrapolating a TRGB slope calibration from different fiducial colors.}

\section{Discussion}

{In the above sections, the Cepheids, JAGB, and TRGB were each identified from JWST imaging of NGC~4536. The TRGB's color dependence was then examined in closer detail through various analyses. In this section we place the TRGB slope findings in context with the literature and discuss the prospects for future TRGB measurements with JWST.}

\subsection{TRGB Results from This Study}

{The color slope of the IR TRGB in the considered JWST bands was estimated by tracing tip stars identified in overlapping HST imaging to their cross-matched NIRCAM magnitudes, then performing various minimizations--the intent being broad coverage of different approaches to treatment of measurement uncertainties. The estimated slopes ranged from $-0.50$ to $-1.12$~mag/mag. Shallower values came from conventional least squares approaches and steeper ones from accounting for the larger uncertainties on the color axis. The ODR slope of $-0.99 \pm 0.16$~mag/mag should ostensibly be the most accurate inference for uncertainties along both axes, given accurate measurement uncertainties and identification of TRGB stars.
}

{Shown in \autoref{fig:rect_trgb_measure}, this slope value was then used to rotate the halo-selected CMD and flatten the TRGB feature (or ``rectification''). A one-dimensional edge detection was computed from the RGB LFs constructed from each of the unrectified and rectified CMDs. The EDR computed from the unrectified RGB LF showed a clear bimodality, with each peak falling at the blue and red ends of the sloped TRGB feature. The EDR of the rectified RGB LF becomes singularly-peaked, indicating most of the slant of the TRGB in the CMD was successfully removed.}

{The separation of the two peaks in the unrectified bimodal EDR was 0.28~mag, while the width of the rectified single-peaked EDR was 0.09~mag. \citet{McQuinn_2019} briefly mention the F115W band in their isochrone overview paper and point out that the variation in TRGB luminosity due to metallicity and age is 0.3 and 0.09~mag, respectively. This is consistent with the hypothesis that our slope has corrected for metallicity effects in the observed TRGB, and that a sizable portion of the remaining dispersion in the slope-corrected measurement is possibly sourced by uncorrected age effects. Though the current number statistics preclude any further or more rigorous interpretation.}

{The impact of the color slope correction on TRGB measurement was then explored over a finer grid of rectification slope values. The results were illustrated in \autoref{fig:running_slope_edrs} and the information condensed into the 68\% width of the TRGB feature in \autoref{fig:running_widths}. The results revealed that $\beta$ values in the range $-0.7$ to $-1.5$ mag/mag minimized the width of the TRGB feature, which is consistent with the slope values inferred earlier via the optical-IR trace method. It is important to note that the majority of the imaging used in these tests was \textit{not} contained in the HST+JWST overlap region that was used to do the optical-IR tracing. See the hatched (HST+JWST) vs. unhatched (JWST only) regions in \autoref{fig:pointings}.
}

{The width of the TRGB feature discussed in the previous paragraph can be viewed as a proxy for the statistical uncertainty, or the precision of a \textit{relative} distance. The results demonstrated the improvement in this precision that is attained when applying a correction for the TRGB color slope, and that there exist a range of slope values that best maximize the statistical precision. Now we discuss the effect that the TRGB slope (and its uncertainty) would have on an \textit{absolute} TRGB distance measured relative to a fiducial zeropoint calibration.}

{In \autoref{fig:running_delta_mu_fidu}, the running slope experiment was reformulated to explore the measurement of an absolute distance to this (or any) target galaxy. The shift in estimated distance was plotted as a function of $\beta$ and at four values of $c_0$. Of course, the curve was flat when the target CMD was rotated about the mean color of the target TRGB, i.e., the fiducial was set to itself. Then, three hypothetical color fiducials were also plotted, to demonstrate two parts of the distance uncertainty related to TRGB color slope correction: 1) the propagation of slope uncertainties into distance uncertainty, and 2) the amplification of that uncertainty as the difference between the color of the fiducial calibration relative to the target galaxy increases.}
{As expected, and shown in \autoref{fig:running_delta_mu_fidu}, the deviations in estimated distance are larger for fiducial calibrations that are further offset from the color of the TRGB in the target galaxy. That is, for the same TRGB slope values, $\Delta \mu$ is larger for the curve based on a $c_0 = 1.51$~mag fiducial than it is for the one based on a $c_0 = 1.41$~mag fiducial.}

{We can also view this effect from a more practical perspective with some approximate calculations. The slope inferred from the ODR fit to the optical-identified tip stars was $\ODRval \pm \ODRerr$~mag/mag. If one's fiducial TRGB calibration has the same $c_0$ as the target galaxy, then the additional uncertainty from the slope extrapolation would be zero. However, if the fiducial color is 0.2~mag offset in $c_0$, then the TRGB slope uncertainty propagated into the distance uncertainty would be equal to $0.2\mathrm{mag} \times 0.16\mathrm{mag/mag} = 0.032$~mag. This uncertainty can also be visualized in \autoref{fig:running_delta_mu_fidu} as the vertical distance between the two points where the $c_0 = 1.51$~mag fiducial (green curve) intersects the projection of the ODR slope and error (red box).}

{This demonstrates that the color of one's fiducial TRGB calibration should be as close as possible to the color of the target galaxy's TRGB in order to minimize slope extrapolation uncertainties. This effect is identical to the systematic error observed in applications of the Cepheid Leavitt Law when there is an offset between the mean period of one's calibrator Cepheids and the mean period of a target galaxy's Cepheids.}

{In all cases of our analysis, the slope intervals that best-represented the empirical data contained the values of the TRGB's metallicity slope predicted by the BaSTI isochrone suite for ages 4 and 10~Gyr of a TRGB population, equal to $-0.74$ and $-0.92$~mag/mag, respectively.}

\subsection{Literature TRGB Slope Estimates}
{At the time of writing, there does not exist a detailed study of the TRGB slope in the specific JWST bands discussed here. So we discuss our results in the context of the similar ground-based $J$ and $K$ bands. The effective wavelengths of F115W and $J$ are $1.15~\mu\mathrm{m}$ and $1.24~\mu\mathrm{m}$, respectively. For F444W and $K$ they are $4.4~\mu\mathrm{m}$ and $2.2~\mu\mathrm{m}$, respectively. The bandpasses in the blue pair are separated by only 90~nm in effective wavelength, while the wavelength coverage of both the redder bandpasses is well into the Rayleigh-Jeans tail of an RGB star. This can be verified in Fig. 5 of \citet{McQuinn_2019} wherein the K, F277W, [3.6], and [4.5] bands all have the same TRGB magnitude on the Vega system. As a result, the $JK$ bands should provide a reasonable approximation of the F115W, F444W bands.}

{The slopes we have derived here, from either the optical-identification approach (i.e., \autoref{tab:slopes}) or the running-slope experiment (\autoref{fig:running_widths}), are consistent with literature determinations in these similar ground-based bandpasses. \citet{Valenti_2004} reported for $J$ vs. $(J-K)$ a slope of $-1.15$~mag/mag from a sample of globular clusters.\footnote{The \citeauthor{Valenti_2004} calibration is presented in terms of $[Fe/H]$. For a transformation of their $[Fe/H]$ equations to the $J,JK$ ones considered here, see \citet{Hoyt_2018}.} \citet{Madore_2018} found $-0.85 \pm 0.12$~mag/mag from a study based in the Local Group galaxy IC~1613. And \citet{Serenelli_2017} found $-0.81$~mag/mag from a comprehensive study of stellar model predictions, testing variations in input physics as well as bolometric corrections, ranging from empirical measurements to theoretical atmosphere models.\footnote{The \citeauthor{Serenelli_2017} $M_J$ vs. $(J-K)$ calibration is presented as a second order equation, but the linear term is an order of magnitude more dominant for $(J-K) > 0.90$~mag, which is about the color of SMC TRGB stars, so we propagate only the linear term here.}} 

\subsection{Broader Discussion on Viability of IR TRGB}
{The viability of the IR TRGB as a precision distance indicator has been questioned at times over the last two decades, largely because of its significant color/metallicity dependence. In particular, \citet{Gorski_2016} used a TRGB calibration based on Galactic globular clusters \citep{Bellazzini_2001, Valenti_2004} to estimate the distances to the Magellanic Clouds, and concluded that the TRGB was inconsistent at the 0.4~mag level, based on the discrepancy between their $I$ and $JHK$ estimations of the distance to each Cloud.}
{However, the authors chose to adopt spectroscopic metallicities that were derived from each cloud's inner-most stars (not necessarily old RGB stars) as representative of the metallicities of their TRGB stars. This may have led to an overestimate of the metallicity of the LMC TRGB stars of order 0.2~dex. For instance, APOGEE mapped abundances to CMD-selected RGB stars across the face of the LMC. The typical values of [Fe/H] ranged from -1.0 to -0.5~dex, with a long tail towards Fe-depleted and a sharp truncation at the metal-rich end \citep{Nidever_2020}. The metallicity of [Fe/H] = -0.6~dex adopted by \citeauthor{Gorski_2016} lies near the upper-most end of the APOGEE distribution.}

{If we instead assume a value of -0.75~dex for the LMC TRGB stars, which is closer to the median of the APOGEE distribution, the \citeauthor{Gorski_2016} $I$, $J$, and $K$ distance moduli would shift from 18.29, 18.63, and 18.70~mag to 18.37, 18.59, and 18.63~mag, respectively. All the adjusted moduli are much closer to the detached eclipsing binary distance of 18.48~mag \citep{Pietrzynski_2019} than the original values. The fact that a simple shift to the mean metallicity would improve the distances they derived from \textit{both} the $I$ and the $JK$ TRGB, which have oppositely signed metallicity slopes, provides additional confidence to this hypothesis of overestimated metallicities to TRGB stars. This view is also consistent with the discussion already covered by \citeauthor{Gorski_2016} on their adopted spectroscopic vs. RGB-color-inferred metallicities. Most of the remaining disagreement is likely due to discrepancies in photometric zero points between the \citet{Valenti_2004} globular cluster catalogs and IR catalog of the LMC adopted by \citeauthor{Gorski_2016}.}

\subsection{A Unique Advantage of the IR TRGB with JWST}
{The JWST/NIRCAM F090W filter is the bluest viable bandpass for making precision TRGB measurements at large distance.\footnote{{There are no advantages to using the F070W band for TRGB. From an instrument perspective, it is a less sensitive band than F090W and has an extremely under-sampled PSF. And from the astrophysics perspective, TRGB stars are much fainter in F070W than in any other band installed on JWST.}}
As a result, any ``secondary'' band that is chosen to construct a CMD will always be redder than the ``primary'' one that was used to derive the TRGB distance.}
{The secondary band being redder than the primary one carries with it some simple, yet immensely beneficial consequences. Contamination from AGB stars is reduced, particularly when using the F115W band as the primary band. And one is able to fully populate the RGB, unlike the red-truncated RGBs seen in optical CMDs such as in $I$ vs. $(V-I)$.
}

{C-rich AGB stars and Ext-AGB stars, which can contaminate the TRGB, are pulled away from the TRGB in the NIR. That is, the fact that the earlier-discussed JAGB population is identifiable on the CMD immediately reduces the odds that a TRGB measurement can be biased by the presence of an AGB population. Such misidentifications have led to significant ($>0.4$~mag) biases in TRGB measurements such as some of the ones presented in \citet{Scolnic_2023}.
}

{In a similar vein, because of the TRGB's negative slope in the IR, the detection of metal-poor TRGB stars immediately guarantees that one has a complete sample of TRGB stars. The loss of metal-rich stars is an observational selection effect imposed when the secondary color is bluer than the primary one. This was highlighted in \autoref{fig:tip_id} in which the most metal-rich (solar metallicity) isochrone fell far off the HST optical CMD, but was fully contained in the JWST IR CMD.}

{In light of these points, the most ideal combinations for doing TRGB with JWST are going to be F090W paired with F150W through F356W, or F115W paired with F200W through F356W. Note the use of F444W as done in this study is certainly possible, but just not ideal since it is the least sensitive NIRCAM bandpass. And the pairing of F115W with F150W provides too small a wavelength baseline, leading to an excessively steep TRGB slope.}

\subsection{Future Improvements}

{We anticipate a significantly better understanding of the IR TRGB slope will come with an enlarged sample of targets observed with JWST. This will lead to decreased statistical and systematic uncertainties in the measurement of TRGB distances with JWST.}

{The formal uncertainty on our slope determination from the optical-IR trace method is $0.16$~mag/mag, or 16\%. And from the running slope experiment, the range of slopes that appeared to flatten what is ostensibly the TRGB feature in the CMD ranged from $-0.7$ to $-1.5$~mag/mag. Improving these constraints will come as JWST continues to observe RGB populations both in our program and in others \citep[e.g.,][]{2021jwst.prop.1638M}.}

{As emphasized in \cite{McQuinn_2019}, the variation in the IR colors and magnitudes of TRGB stars not captured by metallicity is caused by age. It will be important in future work to calibrate the TRGB slope over a range of RGB populations with various star-formation histories, to essentially average over these second-order age variations, and to accurately place one's IR TRGB distance scale onto a universal system that is anchored to a fiducial zero-point (e.g., NGC~4258).}

{Though it should be noted} that environmental variations in both the TRGB's slope and zero-point will likely be minimized when the TRGB is measured in the outer regions of galaxies. This is highlighted in the Appendix, where the TRGB's metallicity-dependent color slope is predicted by the BaSTI suite of stellar evolution models to be only weakly dependent on age in populations above a minimum age threshold ($\sim 6$~Gyr according to those models).

\section{Conclusions}

{In this study, we identified and isolated the TRGB, Cepheids, and JAGB from one set of JWST imaging. The characteristics of all three distance indicators in the observed JWST bandpasses were found to be in excellent agreement with ingoing expectations, including the slope and dispersion of the Cepheid Leavitt Law, the dispersion of the JAGB LF, and the slope of the IR TRGB.}

{The TRGB was then explored in greater detail, in particular its color dependence. First, TRGB stars were identified in overlapping HST imaging and their cross-matched JWST magnitudes used to estimate the TRGB slope.} The TRGB slope parameter was then inferred using several different statistical techniques to account for the lower signal-to-noise of the secondary F444W photometry.

{Several experiments were used to estimate the impact of the TRGB slope on uncertainties in distance measurement. Notably, a bimodal pair of edge features in the uncorrected CMD was merged into a single discontinuity in the rectified CMD. This translated to an approximate reduction in TRGB magnitude measurement uncertainty from about 0.3~mag to a value less than 0.09~mag, demonstrating that the IR TRGB as observed in JWST bandpasses can be used as a precision distance indicator.}

{The IR TRGB slope estimates were consistent with predictions from the BaSTI suite of isochrones for old RGB populations, demonstrating that our understanding of the TRGB's color variation is already off to a promising start.}
Future observations from this and other JWST programs will {bolster the sample of TRGB stars we can use to converge on a more robust estimate of the TRGB's color dependence and subsequently establish a precise and accurate JWST TRGB distance scale in the infrared.}

We have demonstrated here that three entirely independent distance indicators---Cepheid variables, the TRGB, and the Carbon star luminosity function (or JAGB)---can all be simultaneously detected and identified {from imaging in just one JWST pointing.} The next {NASA} flagship looks {more than capable of providing} us with a better understanding of {potentially unknown} systematic uncertainties that may still underlie the measurement of extragalactic distances and, subsequently, {distance ladder} determinations of the Hubble constant.

\section*{Acknowledgements}
\begin{acknowledgments}
TJH acknowledges Saul Perlmutter and Greg Aldering for their support of his research, {as well as for feedback on this manuscript and its contents}. TJH's contributions to this work were supported in part by the U.S. Department of Energy, Office of Science, Office of High Energy Physics under Contract No. DE-AC02-05CH11231. We thank the Stellar Populations Early Release Science (ERS) team (PI: Weisz) and A. Dolphin for their provision of an early release version of their {DOLPHOT software to do} JWST/NIRCam photometry. We acknowledge the usage of the HyperLeda database (\url{http://leda.univ-lyon1.fr}). This research is based on observations made with the NASA/ESA Hubble and the NASA/ESA/CSA James Webb Space Telescopes obtained from the Space Telescope Science Institute, which is operated by the Association of Universities for Research in Astronomy, Inc., under NASA contract NAS 5–26555 for HST, and under NASA contract NAS 5-03127 for JWST. These observations are associated with programs HST-GO-13691 and JWST-GO-1995. {Data were downloaded from the Mikulski Archive for Space Telescopes (MAST) hosted by the Space Telescope Science Institute and can be accessed via \dataset[10.17909/rvxy-pt92]{https://doi.org/10.17909/rvxy-pt92}.}
We thank the University of Chicago and the Observatories of the Carnegie Institution for Science for their past and ongoing support of our long-term research into the calibration and determination of the expansion rate of the Universe. {We thank the anonymous referee for their feedback and suggestions.}
\end{acknowledgments}

\facilities{JWST(NIRCam) HST(ACS/WFC)}

\software{DOLPHOT \citep{Dolphin_2000, Weisz_2024}, astropy \citep{astropy:2013, astropy:2018, astropy:2022}, jwst \citep{Bushouse_JWST_Calibration_Pipeline_2023}, Jupyter Notebook \citep{Kluyver2016jupyter}, scipy \citep{2020SciPy-NMeth}
}

\appendix
\section{BASTI Isochrones and Predicted TRGB Slopes}

\begin{figure}
    \centering
    \includegraphics[width=0.6\textwidth]{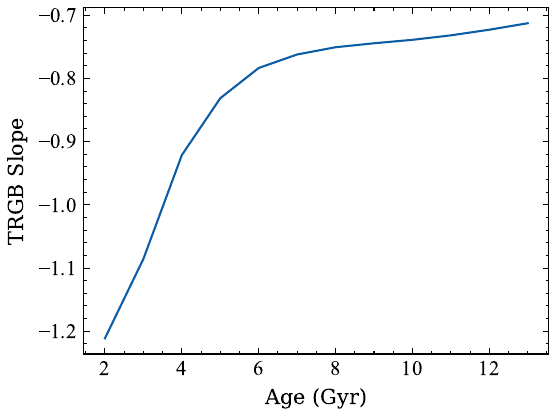}
    \caption{The TRGB's metallicity-dependent slope determined from [Fe/H] = $-1.90$ to $+0.06$ in the F115W vs. $(\mathrm{F115W} - \mathrm{F444W})$ CMD as predicted by the BASTI stellar evolution models for a range of ages that can be expected in galaxy stellar halos. The predicted metallicity-dependent slope decreases and the trend flattens for older populations $t_{age} \gtrsim 6$~Gyr, suggesting that targeting pristine stellar halo populations will be key to reducing population-dependent systematic errors in the measurement of TRGB distances in the IR with JWST.}
    \label{fig:basti_slopes}
\end{figure}

The BASTI database \citep{Hidalgo_2018, Pietrinferni_2021} was queried over a grid of ages from $2000-13000$~Myr in 1000~Myr intervals and for metallicities $[\mathrm{Fe}/\mathrm{H}] = \{-1.90, -1.55, -1.30, -1.05, -0.60, -0.30, +0.06\}$. The brightest magnitude for each isochrone in the F115W band is then selected and used along with its corresponding $(\mathrm{F115W} - \mathrm{F444W})$ color to predict the TRGB color slope as a function of metallicity within each age bin. The result is shown in \autoref{fig:basti_slopes} which demonstrates that the metallicity-dependent slope of the TRGB is predicted to vary less with age for RGB stars beyond a certain minimum age threshold (in this case $\sim 6$~Gyr).

This is promising in that we have reason to believe that by ensuring that we measure the IR TRGB in old-age populations, we can minimize the concomitant increase in dispersion necessarily incurred when transitioning TRGB measurements from the color-insensitive 800~nm to the color-dependent IR. This is imperative to emphasize in light of many recent studies that have attempted to measure the TRGB from regions of galaxies that are contaminated by young and/or intermediate-age populations \citep[e.g.,][]{Reid_2019, Yuan_2019, Anand_2022, Wu_2023, Scolnic_2023}.

\bibliography{main}{}
\bibliographystyle{aasjournal}

\end{document}